\documentstyle[preprint,tighten,aps,amstex,graphicx,times,floats]{revtex}

\newcommand{\eg}{{\it e.g.}\;}
\newcommand{\ie}{{\it i.e.}\;}
\newcommand{\etal}{{\it et al.}\;}

\newcommand{\msbar}{\overline{{\rm MS}}}
\newcommand{\gev}{\; \mbox{Ge$\!$V}}
\newcommand{\pb}{\; \mbox{pb}}

\newcommand{\epc}[3]{${\rm Eur. Phys. J.}$ {C#1} (#2) #3}
\newcommand{\zpc}[3]{${\rm Z. Phys.}$ {C#1} (#2) #3}
\newcommand{\npb}[3]{${\rm Nucl. Phys.}$ {B#1} (#2)~#3}
\newcommand{\plb}[3]{${\rm Phys. Lett.}$ {B#1} (#2) #3}

\renewcommand{\prd}[3]{${\rm Phys. Rev.}$ {D#1} (#2) #3}
\renewcommand{\prl}[3]{${\rm Phys. Rev. Lett.}$ {#1} (#2) #3}

\newcommand{\hep}[1]{${\rm hep\!-\!ph/}${#1}}

\begin{document}

\thispagestyle{empty}

\title{ 
~\vspace*{-2cm} \\
{\normalsize \rm
\begin{flushright}
MAD--PH--00--1185 
\end{flushright} }
Single Stop Production at Hadron Colliders
} 

\author{ T.~Plehn\footnote{Supported in part by DOE grant 
                  DE-FG02-95ER-40896 
                  and in part by the University of Wisconsin Research 
                  Committee with funds granted by the Wisconsin Alumni 
                  Research Foundation.} } 

\address{\vspace*{2ex}
Department of Physics, 
University of Wisconsin, Madison, WI 53706, USA
} 

\maketitle 

\begin{abstract}
  We analyze the production of a single top squark at hadron
  colliders: $p\bar{p}/pp \to \tilde{t}_1 + X$.  The total cross
  sections and the transverse momentum distributions are presented in
  next-to-leading order QCD. The higher-order corrections render the
  predictions theoretically stable with respect to variations of the
  factorization and renormalization scales. The transverse momentum
  distribution is resummed to estimate effects of the small
  transverse momentum regime on possible analyses. Since the
  corrections increase the cross sections and reduce the theoretical
  uncertainty, the discovery range for these particles is extended in
  the refined analysis.
\end{abstract}

\vspace*{5mm}

%%%%%%%%%%%%%%%%%%%%%%%%%%%%%%% MAIN TEXT %%%%%%%%%%%%%%%%%%%%%%%%%%%%
Hadron colliders like the Tevatron and the LHC will soon be able to
either discover or strongly constrain physics at scales well above the
Standard Model masses. From an aesthetic point of view, the most
attractive realization of supersymmetry could be the minimal
supersymmetric model (MSSM). However, $R$ parity conservation is an
{\sl ad hoc} assumption, invoked to bypass problems with
flavor--changing neutral currents, proton decay, atomic parity
violation and other experimental constraints. From a more general
point of view, these observables put tight limits on some, but not on
all, $R$ parity violating couplings. For collider searches, exact $R$
parity predicts that supersymmetric particles can only be
pair-produced. Limited by the beam energy, the Tevatron tends to run
out of supersymmetry discovery reach if the strongly interacting
squarks and gluinos become too heavy to be produced in pairs. But many
models based on unification scenarios at some high scale prefer
exactly this kind of mass spectra.  If, in contrast, $R$ parity is not
an exact symmetry, the signals from single superpartner production can
be nicely extracted in the low background environment at the
Tevatron~\cite{report,bhs,tev_more} or even at the
LHC~\cite{herbi}.\smallskip

In the case of a light scalar top squark, the baryon number violating
coupling $\lambda''_{ijk}$ induces the production process $\bar{d}_j
\bar{d}_k \to \tilde{t}$~\cite{bhs}. It stems from the superpotential
contribution~\cite{weinberg}
\begin{equation}
{\cal W} = \lambda_{ijk}'' U_i^c D_j^c D_k^c,
\end{equation} 
where $D^c$ and $U^c$ denote charge conjugate right handed quark
superfields. The couplings including at least one third generation
flavor index $\{i,j,k\}$ are currently only constrained in the
combination $|\lambda_{313}''
{\lambda_{323}''}^*|$~\cite{constraints,kmass}. The relevant
Lagrangean reads\footnote{Our conventions follow the leading order
  signal and background analysis~\cite{bhs}.}
\begin{equation}
{\cal L}_{\lambda''} = -2 \varepsilon^{\alpha \beta \gamma}
                        \lambda_{3jk}'' 
                        \left[ \tilde{t}_{R \alpha}
                               \overline{d_{j \beta}^c}
                               P_R d_{k \gamma} + {\rm h.c.}
                        \right] + \cdots
\qquad \qquad (j<k)
\end{equation}

If we assume that the $R$ parity violating coupling is a small
parameter $\lambda'' \lesssim 0.1$, and the stop is not too light,
the produced single top squark has a sizeable branching fraction into
the 'classical' supersymmetric channel $\tilde{t} \to b
\tilde{\chi}^+$, where the light chargino decays to $l \nu
\tilde{\chi}_1^0$~\cite{bhs}. These decays have been calculated in
next-to-leading order~\cite{stop-decay}; as expected for a decay
mediated by the weak coupling constant, the corrections is moderate
($K \lesssim 1.2$), and the scale dependence is well under control.
Furthermore, the higher order corrections leave the dependence on the
stop mixing angle essentially unchanged.\bigskip

\begin{figure}[t] 
\begin{center} \vspace*{-4mm}
\includegraphics[width=17.5cm]{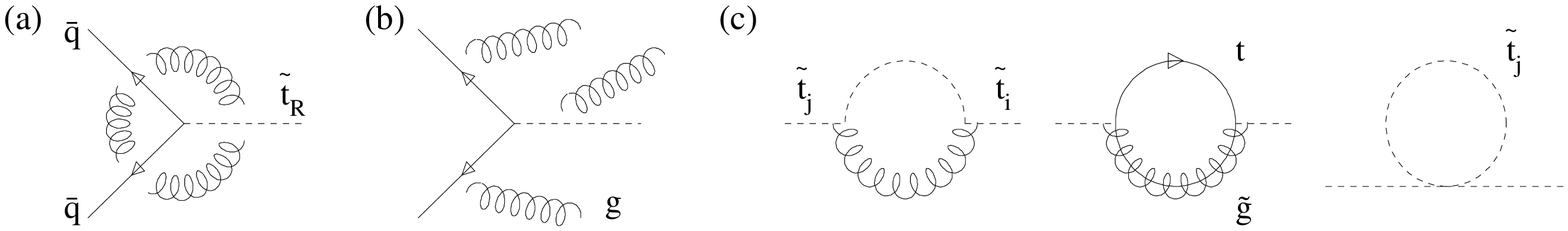}
\end{center} \vspace*{-4mm}
\caption[]{\label{fig:feyn} {\sl 
  Basic diagrams for the single stop production at hadron colliders in
  antiquark-antiquark collisions: (a) generic diagrams of virtual QCD
  corrections; (b) generic diagrams for real gluon emission; (c)
  SUSY-QCD corrections to the stop propagator, including the 
  non-Standard Model contributions to the running mixing angle.}}
\end{figure}

\noindent 
{\bf \underline{Next-to-Leading Order Cross Section}}
\\[2mm] 
In leading order the partonic cross section for the production of a
single light top squark $\tilde{t}_1$ with the mass $m$ is given by
\begin{equation}
\hat{\sigma}_{\rm LO} = K_{qq} \;
                        \frac{4 \pi N_c (N_c-1)}{m^2} 
                        \sin^2 \theta_t \; |\lambda''|^2 \;
                        \delta\left( 1 - \frac{m^2}{s} \right)
                      \equiv K_{qq} \;
                      \frac{\sigma_0}{m^2} \;
                      \delta\left( 1 - \frac{m^2}{s} \right) ;
\qquad \quad
K_{qq} = \frac{1}{4 N_c^2},
\label{eq:lo}
\end{equation}
where the stop mixing angle $\theta_t$ measures the fraction of the
right handed squark field in the light mixed state. During the
following analysis we will omit this over-all factor, \ie we assume
$\tilde{t}_1 = \tilde{t}_R$. All cross sections can be scaled
na\"{\i}vely; we present the details of this argument below.\smallskip

\begin{figure}[b] 
\begin{center} \vspace*{-4mm}
\includegraphics[width=7.8cm]{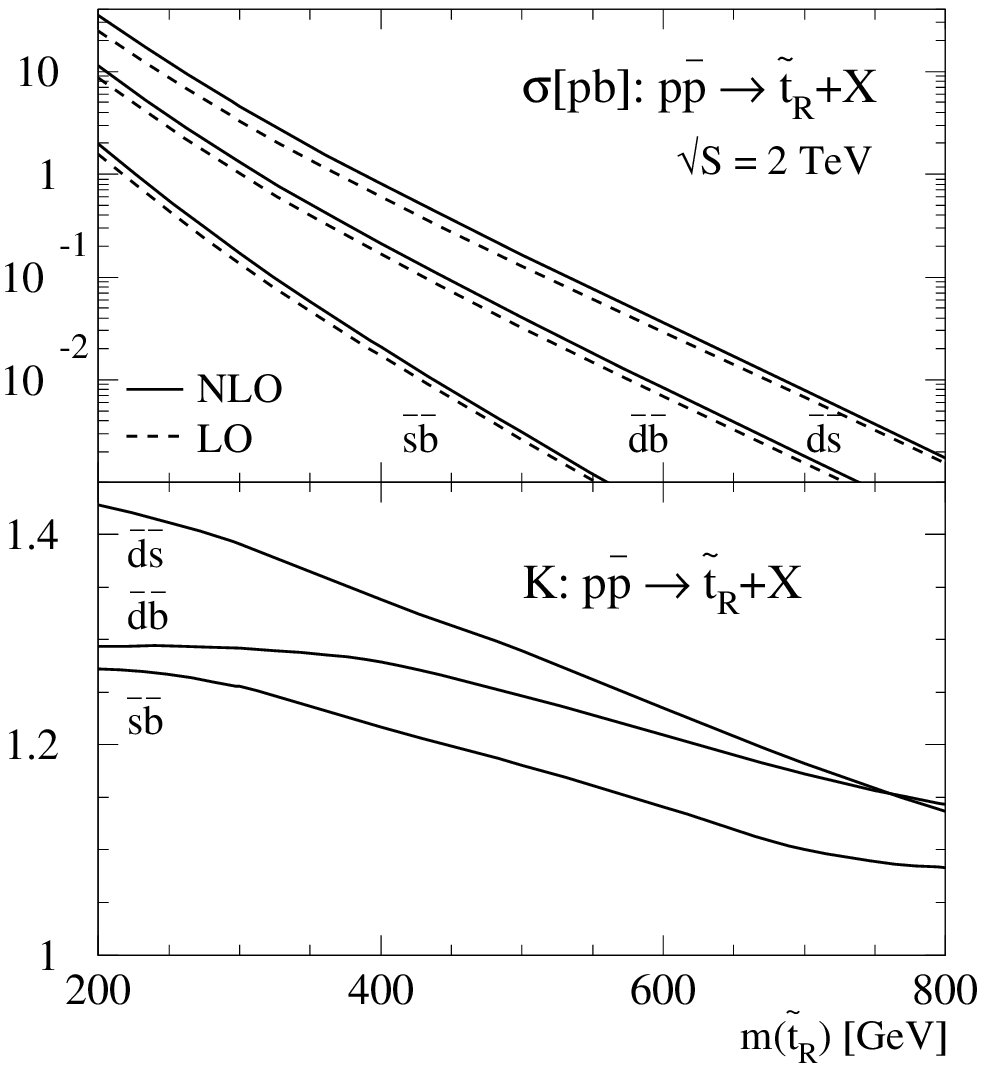} \hspace{1cm}
\includegraphics[width=7.8cm]{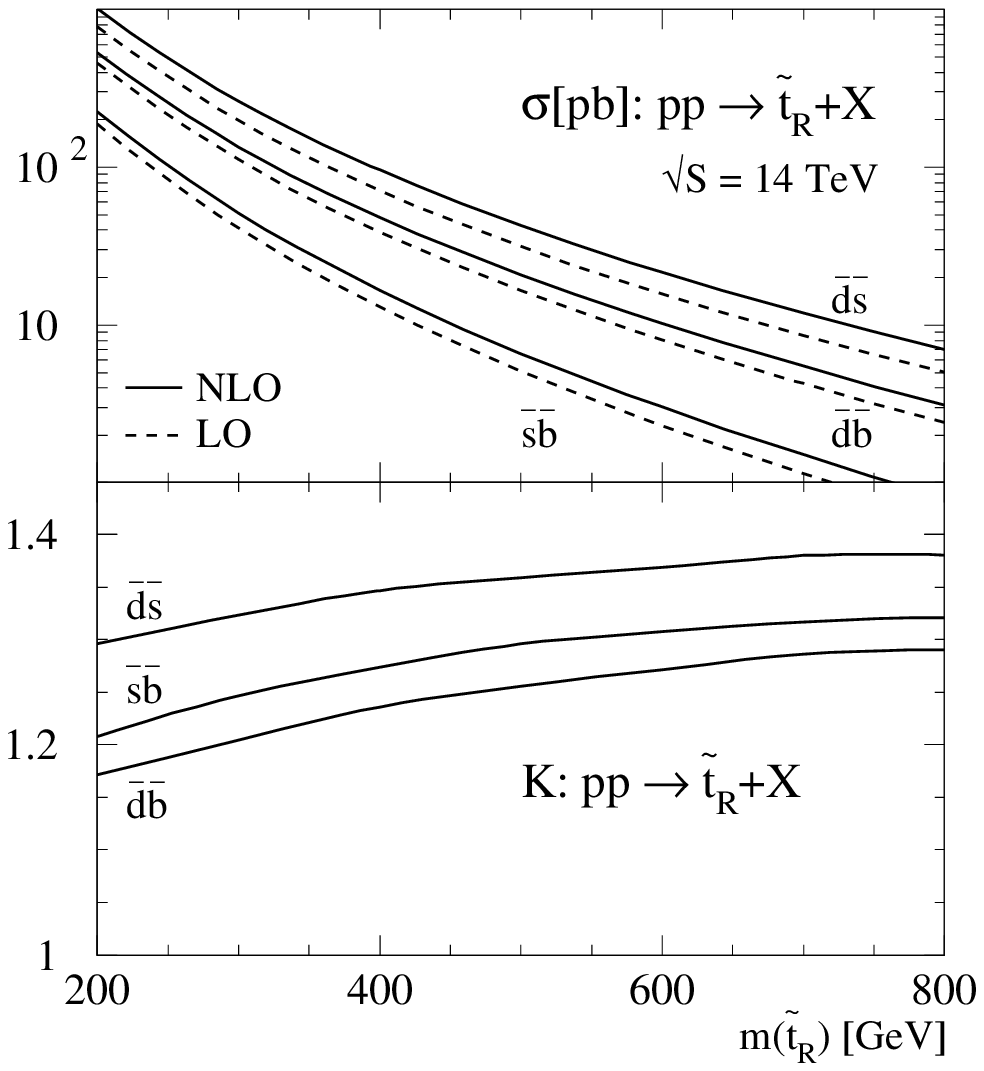}
\end{center} \vspace*{-4mm}
\caption[]{\label{fig:kfac} {\sl 
  Total cross section and $K$ factors for the single $\tilde{t}_R$
  production as a function of the mass. The couplings $\lambda''$ are
  all set to 0.1. The mixing angle dependence is omitted, $\sin
  \theta_t=1$; the renormalization and factorization scales are fixed
  to the final state mass, and the parton densities used are
  CTEQ5L/M1~\cite{cteq}.  The curves are calculated for $\tilde{t}$
  plus $\tilde{t}^*$ production for the three different couplings
  $\lambda''_{ijk}$. The dashed curves show the leading order, the
  solid curve the next-to-leading order results.}}
\end{figure}

The ${\cal O}({\lambda''}^2\alpha_s)$ cross section includes virtual
gluon contributions, as well as real gluon emission and the crossed
channel with an incoming gluon (Fig.~\ref{fig:feyn}). The infrared and
ultraviolet divergences are extracted using dimensional
regularization. After adding virtual and real gluon contributions, and
after renormalization, only collinear divergences remain. They are
absorbed into the definition of the parton densities through mass
factorization. Numerically we use the kinematical delta distribution
in the Born cross section, eq(\ref{eq:lo}), to perform the convolution
of splitting functions and the leading-order cross section.

Diagrams involving heavy supersymmetric particles, like stops and
gluinos, contribute to the virtual next-to-leading order matrix
elements. From stop pair production~\cite{stop-pair}, we know that the
contribution of these loops is numerically well below the remaining
theoretical and experimental uncertainties. Therefore we disregard
them in the following analysis to remain maximally independent of the
underlying model. In the case of single stop production there are no
vertex diagrams involving gluinos/heavy stops.\footnote{The coupling
  $dd\tilde{t}$ is in fact related to \eg
  $\tilde{d}\tilde{d}\tilde{t}$ by supersymmetry, but since the latter
  is a three scalar soft breaking parameter we can assume
  $\lambda''_{\tilde{d}\tilde{d}\tilde{t}} \ll
  \lambda''_{dd\tilde{t}}$.}  The supersymmetric partners contribute
through wave function renormalization of external quarks and stop
legs. A second class of corrections arises from the Feynman diagrams
shown in Fig.~\ref{fig:feyn}(c): the external light stop can mix to a
heavy stop, which then couples to the incoming quarks. This is not
part of the wave function renormalization, but it can be absorbed into
a running stop mixing angle $\theta_t(\mu)$, evaluated at the external
particle's mass scale. The Green's function with an external light
stop leg will only depend on the heavy stop mass through the running
mixing angle, \ie the single stop production cross section will be
proportional to $\sin \theta_t$ in leading and in next-to-leading
order. The numerical effect of the running mixing angle has been shown
to be negligible~\cite{stop-decay}.\smallskip

The hadronic $p\bar{p}$ and $pp$ cross sections are obtained by
folding the partonic cross section with parton luminosities. In
leading order only $\bar{q}\bar{q}$ initial states contribute to the
production process. Especially for the production close to threshold,
the gluon luminosity at the Tevatron is expected to be small. At the
LHC, however, the $\bar{q}g \to \tilde{t}+$jet cross section can be
large. To estimate the gluonic contribution, we compute the total
cross section for $\tilde{t}+$jet production cross section
(Tab.~\ref{tb:cxn}). The transverse momentum is cut at a minimum value
of $5 \gev$. A resummed cross section, as it is derived later in this
paper, cannot be used, since in the given order it does not
distinguish clearly between the different incoming states. For a small
stop mass of $200 \gev$, $40\%$ of the $\tilde{t}+$jet events at the
LHC involve an initial state gluon. This fraction drops to $25\%$ for
a stop mass of $500 \gev$. Even at the Tevatron, light stops in
association with a jet are produced through incoming gluons in $25\%$
of the cases.  This large fraction corresponds to the fact that for
$\bar{q}\bar{q}$ initial states the large valence quark luminosity
does not contribute.  \smallskip

The next-to-leading order total cross sections are presented in
Figure~\ref{fig:kfac}. The leading order results are in agreement with
Ref.~\cite{bhs}, after taking into account the large effect from
switching the parton densities from CTEQ4M to CTEQ5M1. The
next-to-leading order cross sections are parameterized by the factor
$K=\sigma_{\rm NLO}/\sigma_{\rm LO}$.  From the point of view of
parton luminosities, both colliders are very similar; the initial
state $\bar{d}\bar{s}$ involves one valence and one sea quark, after
summing over stop and anti-stop production. Therefore, the fraction of
sea quarks and gluons in the proton dominates the $K$ factor. For
larger stop masses, the gluon luminosity at the Tevatron decreases
rapidly; therefore the $K$ factor drops to $1.15$. At the LHC the
change is less dramatic. For the $\bar{d}\bar{s}$ initial state the
$K$ factor starts from a slightly smaller value than at the Tevatron
and slowly increases from 1.3 to 1.4 with increasing mass. \smallskip

\begin{figure}[t] 
\begin{center} \vspace*{-4mm}
\includegraphics[width=7.8cm]{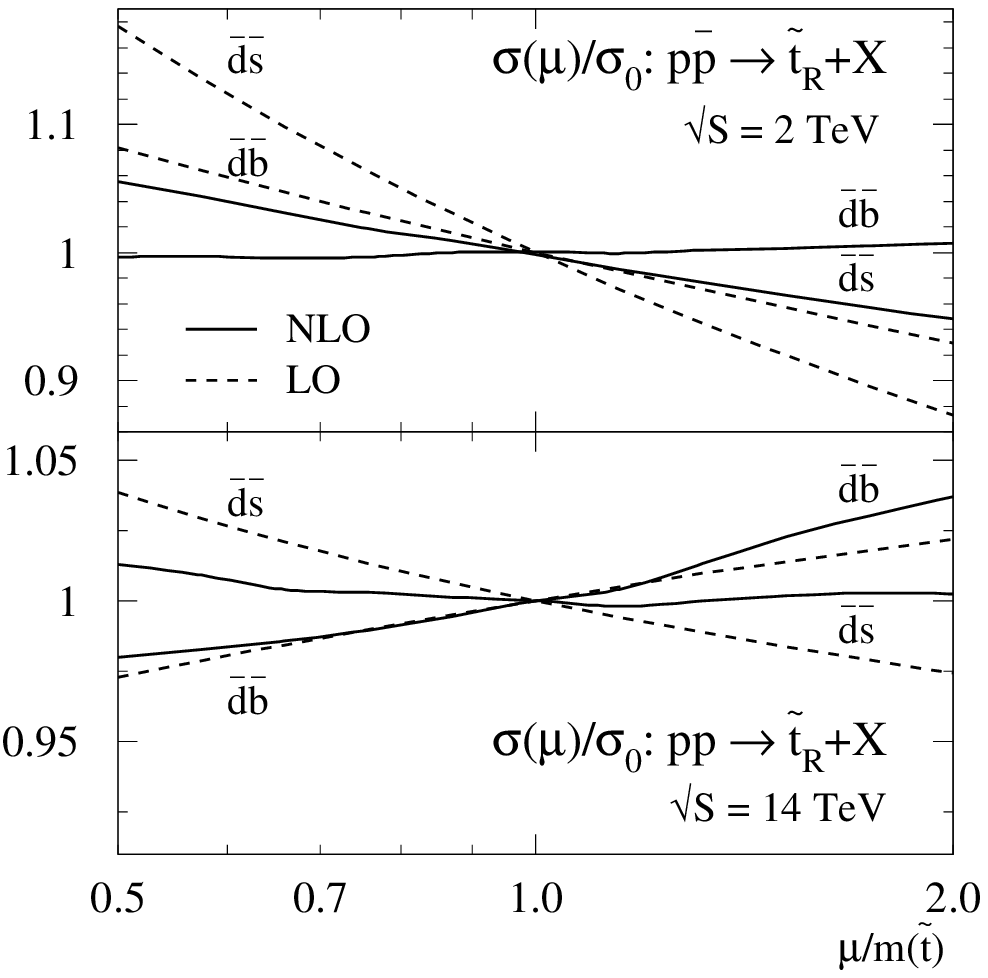} \hspace{1cm}
\includegraphics[width=7.8cm]{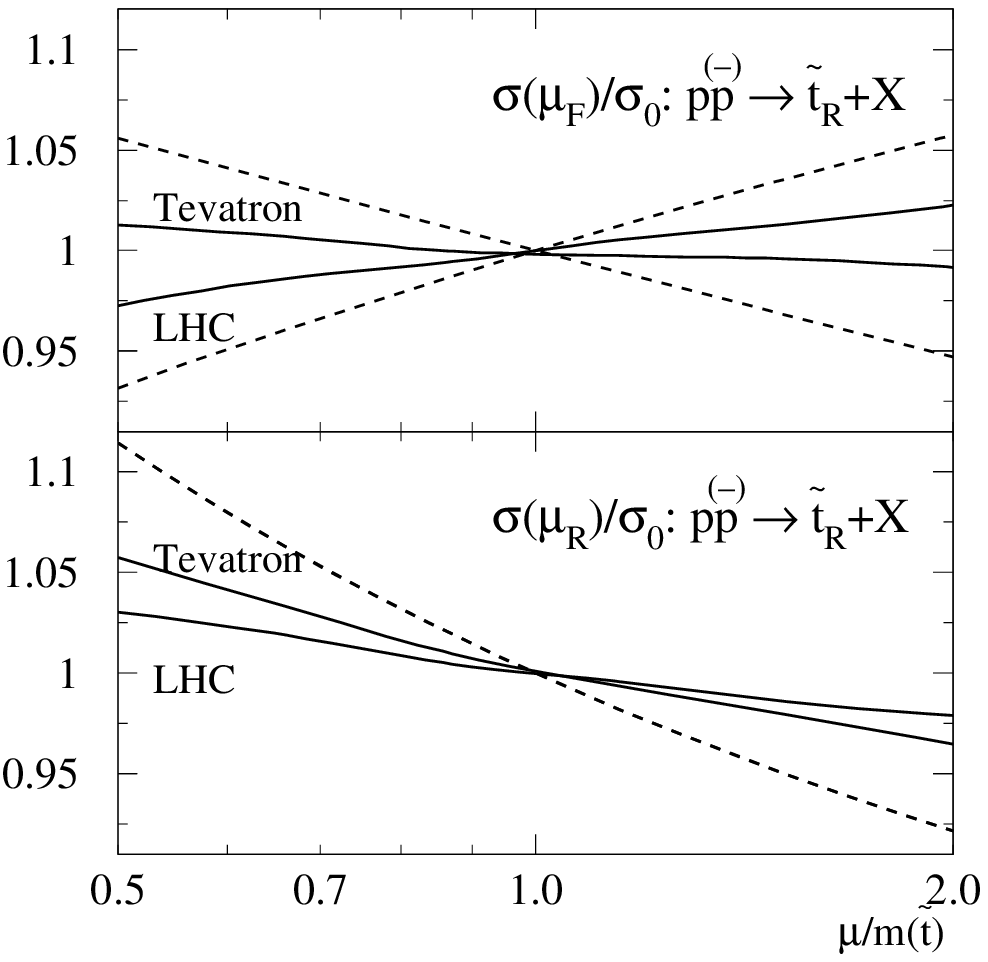}
\end{center} \vspace*{-4mm} 
\caption[]{\label{fig:scale} {\sl 
  Renormalization and factorization scale dependence of the total
  cross section for a stop mass of 200\gev, normalized to the central
  scale value $\mu=m$. Left: both scales are varied for two different
  initial states. The $\bar{s}\bar{b}$ case looks very similar to the
  $\bar{d}\bar{b}$ curve. Right: factorization and renormalization
  scale are varied independently.  The respective other scale is fixed
  to the final state mass. The set of LO and NLO curves are given for
  the Tevatron and the LHC. Note that the leading order
  renormalization scale dependence at both colliders is identical. All
  unspecified parameters are chosen as in Fig.\ref{fig:kfac}.}}
\end{figure}

\begin{table}[b] 
\begin{center} \vspace*{-4mm}
\begin{tabular}{cc||r|r|r|r|r}
 & 
 $m[\gev]$ & 
 $\sigma_{\tilde{t}}^{\rm(LO,CTEQ5L)}$ & 
 $\sigma_{\tilde{t}}^{\rm(LO,CTEQ5M1)}$ & 
 $\sigma_{\tilde{t}}^{\rm(NLO,CTEQ5M1)}$ & 
 $\sigma_{\tilde{t}+g}^{\rm(CTEQ5M1)}$ & 
 $\sigma_{\tilde{t}+q}^{\rm(CTEQ5M1)}$ \\[1mm] \hline 
 Tevatron  & 200 & 24.70\pb & 26.68\pb & 35.28\pb & 11.1\pb  & 3.81\pb   \\
 Tevatron  & 500 & 0.128\pb & 0.113\pb & 0.165\pb & 0.057\pb & 0.0099\pb \\
 LHC       & 200 & 781\pb   & 865\pb   & 1013\pb  & 472\pb   & 321\pb    \\
 LHC       & 500 & 31.4\pb  & 35.7\pb  & 42.8\pb  & 29.2\pb  & 10.9 \pb  \\
\end{tabular}
\end{center} \vspace*{-4mm}
\caption[]{\label{tb:cxn} {\sl
  The total cross sections for single stop production alone, together
  with a gluon jet, and through the crossed gluonic initial state
  together with a quark jet. For the latter the perturbative cross
  section has been used, cut at a minimum value of $5 \gev$. All
  parameters are set as in Fig.~\ref{fig:kfac}.}}
\end{table}

One measure for the theoretical uncertainty coming from higher order
corrections is the dependence of the cross sections on the
renormalization and factorization scales. As a central value for both
scales we use the final state mass ($\mu_F=\mu_R=m$). The dependence
of the total cross section on the scales can be read off
Fig.~\ref{fig:scale}: identifying both scales and varying them between
$m/2$ and $2m$ shows how the uncertainty is reduced by the
next-to-leading order prediction. In case of a $\bar{d}\bar{s}$
initial state the uncertainty drops from $15\%$ to $5\%$ at the
Tevatron and from $5\%$ to even smaller values at the LHC.  However,
part of this unusually small scale dependence in leading and
next-to-leading order is due to a cancellation between the
factorization and the renormalization scale dependences. Especially at
the LHC, the total scale dependence cancels out almost perfectly.
Varying the scales independently yields an uncertainty of $10\%$ in
leading and $5\%$ in next-to-leading order.\bigskip

\noindent 
{\bf \underline{The Coupling $\mathbf{\lambda''}$ in Next-to-Leading
    Order}}
\\[2mm]
The $R$ parity violating coupling $dd\tilde{t}_R$ is not an exclusive
feature of the MSSM. The underlying model could as well be the
Standard Model with an additional strongly interacting scalar; the
coupling $\lambda''$ has no symmetry operation that connects it to any
Standard Model coupling. The MSSM, however,
adds new particles to the spectrum: the light-flavor squark, the
left--handed scalar top quark, and the gluino can contribute to the
next-to-leading order matrix element. For the process considered, the
simple 'SM+$\tilde{t}_R$' model can be regarded as the low energy
limit of the MSSM, after decoupling of the light-flavor squark, the
heavy stop state and the gluino. In this case the diagrams involving
any of these particles vanish; the projector on the right--handed stop
state ($\sin \theta_t$) becomes unity or zero, dependent on if the
lighter stop is the partner of the left or the right--handed top quark.
Since the $\lambda''$ couples strongly interacting particles, it has
to be renormalized~\cite{hera}. Starting out from the known
renormalization constants with and without heavy supersymmetric
particles
\begin{equation}
Z_2^{(q)} = 1 - \frac{\alpha_s C_F}{4 \pi} \; 
                \frac{1+1_{\rm MSSM}}{\tilde{\epsilon}};
\qquad \quad
Z_2^{(\tilde{t})} = 1 + \frac{\alpha_s C_F}{2 \pi} \; 
                        \frac{1-1_{\rm MSSM}}{\tilde{\epsilon}};
\qquad \quad
Z_1^{(dd\tilde{t})} = 1 - \frac{3 \alpha_s C_F}{4 \pi} \; 
                        \frac{1}{\tilde{\epsilon}},
\end{equation}
the relation between the four renormalization constants for the vertex
leads to a renormalization and subsequently a running of the $R$ parity
violating coupling $\lambda''$. Since the heavy particles decouple at
typical hadron collider scales, the coefficient of $\lambda''(\mu_R^2)$
is determined by the low energy effective model:
\begin{equation}
Z_{\lambda''} = 1 - \frac{3 \alpha_s C_F}{4 \pi} \; 
                        \frac{1}{\tilde{\epsilon}} ;
\qquad \qquad \qquad 
\lambda''(\mu_R^2) = \frac{\lambda''(Q^2)}
                       {1 + \frac{3 \alpha_s C_F}{4 \pi}
                            \log \frac{\mu_R^2}{Q^2}}
\end{equation}
The pole in the standard $\msbar$ conventions is given as
$\epsilon/\tilde{\epsilon} = (4\pi)^\epsilon \,
\Gamma(1-\epsilon)/\Gamma(1-2\epsilon)$. This treatment is analogous
to the decoupling of heavy particles from the running of the strong
coupling constant $\alpha_s(\mu_R)$. \bigskip

\noindent 
{\bf \underline{The Cross Section for Small Transverse Momenta}}
\\[2mm]
The perturbative calculation of the next-to-leading order cross
section as presented in the previous section needs to be modified to
investigate the $p_T$ distribution of the top squark. If the
transverse momentum becomes much smaller than the other scales in the
process, \ie the mass of the produced particle, the convergence of the
perturbation series in $\alpha_s$ deteriorates. The series in
$\alpha_s$ has to be replaced by a series in $\alpha_s \log^2
Q^2/p_T^2$~\cite{resum-basics,resum-nonpert,resum-new}. It is possible
to resum all terms at least as singular as $p_T^{-1}$ in the
differential cross section $d \sigma/d p_T$: the formalism has been
successfully applied to Drell-Yan, single vector boson and vector
boson pair, heavy quark, and single Higgs boson
production~\cite{resum-all,resum-series,resum-strong}. We start from
the leading order asymptotic cross section in the limit of small
transverse momentum:
\begin{equation}
\frac{1}{\sigma_0} \; \frac{d\sigma^{\rm asym}}{dp_T dy} = 
K_{qq} \frac{\alpha_s}{\pi} \; \frac{1}{S p_T}
\left[ \left( A \log\frac{m^2}{p_T^2} + B \right)
       f_{\bar{q}}(x_1^0) f_{\bar{q}}(x_2^0)               
       + \sum_{i=\bar{q},g} (f_i \circ P_{\bar{q}\leftarrow i})(x_1^0)
                          f_{\bar{q}}(x_2^0) \right]
       + \left( x_1^0 \leftrightarrow x_2^0 \right)        
\end{equation}
Here $x_{1,2}^0 =e^{\pm y} m/\sqrt{S}$ are the momentum fractions in
the limit of small transverse momentum. The leading order coefficients
for the single stop production can be extracted from the soft and
collinear divergences in the perturbative result. In contrast to the
heavy quark production case~\cite{resum-strong}, they do not consist of
a Drell-Yan type contribution and additional final state radiation
terms. Since the coupling $\lambda''$ violates baryon number there is
no analogue without final state radiation, \ie without strong
interaction charge in the final state. The leading divergence in
$p_T^{-1}$ is given by $A=2 C_F$ and $B=-4 C_F$.\smallskip

The resummed differential cross section involves the integration over
the impact parameter space of a Sudakov-like form factor:
\begin{alignat}{7}
\frac{1}{\sigma_0} \; \frac{d\sigma^{\rm resum}}{dp_T dy} &= 
K_{qq} \; \frac{2 p_T}{S}
\int_0^{\infty} db \frac{b}{2} J_0(b p_T) W(b) \notag \\
W(b) &= 
\; \exp \left[ - \int_{b_0^2/b^2}^{m^2} 
\frac{dq^2}{q^2} \; \frac{\alpha_s(q^2)}{2\pi}
\left( A \log\frac{m^2}{p_T^2} + B \right) \right] \;
f_{\bar{q}}(x_1^0) f_{\bar{q}}(x_2^0)
+ \left( x_1^0 \leftrightarrow x_2^0 \right)
\end{alignat}
The integral in the exponent can be calculated
analytically\footnote{This requires a truncation of the running of the
  strong coupling constant to leading order. We have fixed the leading
  order $\Lambda_{\rm QCD}$ to reproduce the correct value for
  $\alpha_s(M_Z)$; this choice introduces a numerical uncertainty,
  since the average renormalization scale in the process might be
  considerably larger, \eg for a stop mass of $500 \gev$.  Moreover,
  the fitted value of $\Lambda_{\rm QCD}$ for the CTEQ5M1 parton
  densities assumes a next-to-leading order running. Changing the QCD
  scale in the given limits yields a numerical variation of $\lesssim
  5\%$ for the resummed cross section.}; the boundary is canonically
chosen as $b_0=2 e^{-\gamma_{\rm E}}$. The Bessel function $J_0(x)$
oscillates, and the amplitude numerically decreases only slowly for
large transverse momenta. In this regime we use the expansion of the
impact parameter integral for large $p_T$ as compared to the lower
limit of $b$ in the Bessel period considered~\cite{resum-series}.  The
factorization scale for the resummation has been chosen as the mass of
the final state particle, and the parton densities are evaluated at
the point $\mu=b_0/b$.  However, it has been shown that the resummed
cross section is ill-defined for $b > \Lambda_{\rm QCD}$; we follow
one way of parameterizing nonperturbative physics by substituting the
form factor $W(b)$ by~\cite{resum-nonpert}
\begin{equation}
W(b) \to W(b_*) \; 
 \exp \left[ -b^2 g_1 - b^2 g_2 \log \frac{b_{\rm max} m}{2} 
      \right];
\qquad \qquad b_* = \frac{b}{\sqrt{ 1 + b^2/b_{\rm max}^2}}
\end{equation}
The parameters $g_{1,2}$ can be fitted to Drell-Yan
data~\cite{resum-all}, yielding $g_1=0.15 \gev^2$ and $g_2=0.4
\gev^2$.  The cutoff scale for the nonperturbative physics we choose
as $b_{\rm max}=(2 \gev)^{-1}$. The numerical dependence on these
parameters has been shown to be negligible~\cite{resum-all}.  It is
possible to match the large and small transverse momentum formulae
using a matching function
\begin{equation}
\frac{d\sigma^{\rm general}}{dp_T dy} = 
\frac{d\sigma^{\rm pert}}{dp_T dy} +
\frac{1}{1+ \left( p_T/p_T^{\rm match}\right)^4} \;
\left[  \frac{d\sigma^{\rm resum}}{dp_T dy} - 
        \frac{d\sigma^{\rm asym}}{dp_T dy} \right]
\end{equation}
where the matching scale $p_T^{\rm match} \sim m/3$ is the typical
choice. For large transverse momentum the asymptotic and the resummed
cross sections give arbitrary unphysical results; numerically their
contribution to the general cross section is faded out by the matching
function. For small transverse momentum the perturbative and the
asymptotic function approach each other rapidly.  The matching
function smoothes out the behavior of the general solution in the
transition region. The uncertainty induced by the particular choice of
a matching function is only ${\cal
  O}(\alpha_s^2)$~\cite{resum-all}.\smallskip

\begin{figure}[t] 
\begin{center} \vspace*{-4mm}
\includegraphics[width=7.8cm]{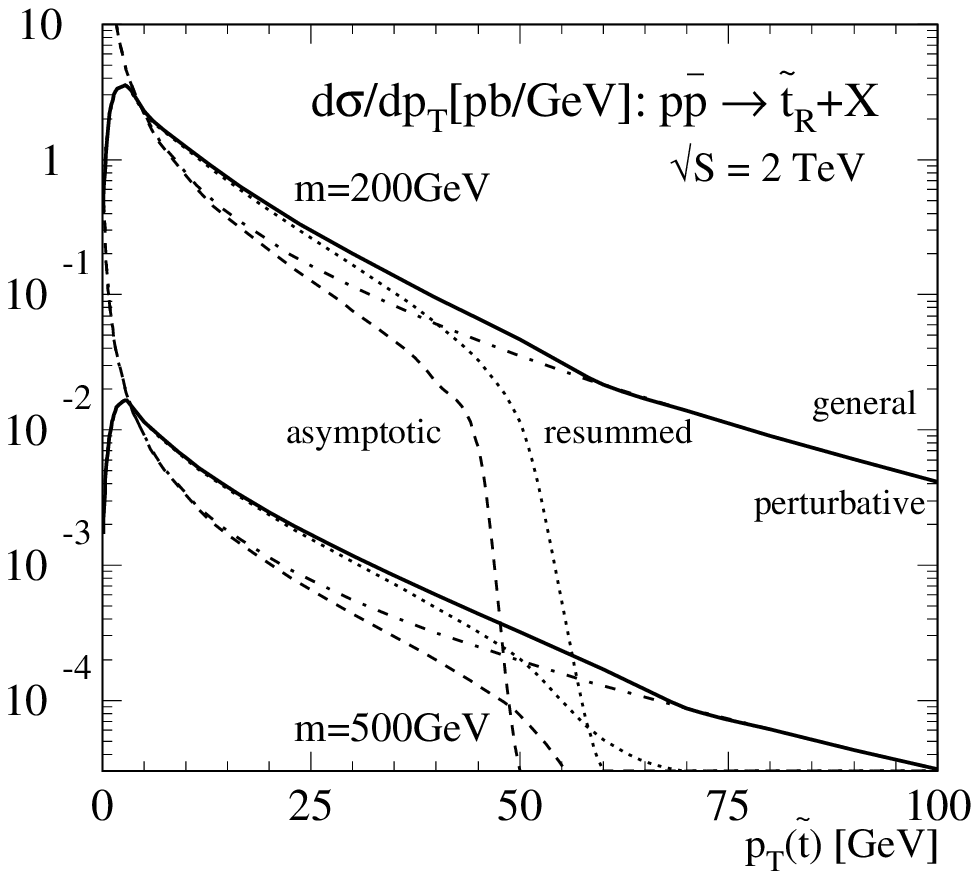} \hspace{1cm}
\includegraphics[width=7.8cm]{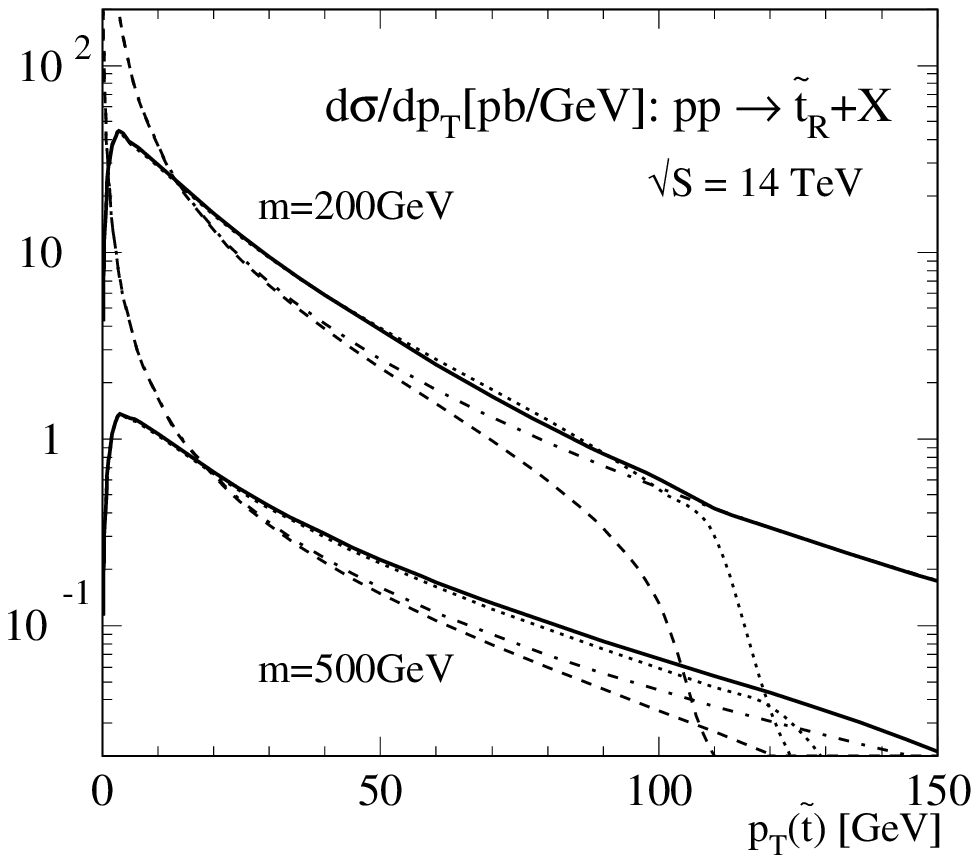}
\end{center}  \vspace*{-4mm}
\caption[]{\label{fig:resum} {\sl
    Differential transverse momentum distribution with respect to the
    top squark. The four different line styles denote the perturbative
    (dot dashed), asymptotic (dashed), resummed (dotted) and general
    (solid) cross sections, as described in the text. The two sets of
    curves correspond to stop masses of 200 and 500\gev. All other
    parameters are chosen as in Fig.\ref{fig:kfac}.}}
\end{figure}

The results of the resummation are presented in
Figure~\ref{fig:resum}. As expected from similar
processes~\cite{resum-all} the resummed cross section peaks at $p_T
\sim 5\gev$. The perturbative and the asymptotic cross sections
numerically agree very well for small transverse momenta $p_T \lesssim
2\gev$. Qualitatively the perturbative and the resummed cross
sections agree for intermediate $p_T \gtrsim 10 \gev$. However,
quantitatively the matched and the perturbative cross sections can
differ by up to $40\%$ for $p_T \lesssim 60 \gev$ at the Tevatron and
$p_T \lesssim 100 \gev$ at the LHC. As can be seen in
Fig.~\ref{fig:resum}, the differential cross sections drop sharply
with increasing transverse momentum; the choice of $p_T^{\rm match}$
should be taken into account as a theoretical source of
uncertainty.\bigskip

\noindent 
{\bf \underline{Conclusions}}
\\[2mm]
It has been shown~\cite{bhs} that the search for single top squarks is
a promising channel for the upgraded Tevatron. We have calculated the
next-to-leading order signal and the $\tilde{t}+$jet production cross
section. The total cross sections are enhanced by up to $30\%$ at the
Tevatron and $40\%$ at the LHC. At both colliders, the crossed
gluon-antiquark incoming states are important, since there is no pure
valence quark production mechanism. However, the typical
next-to-leading order features are similar to supersymmetric pair
production processes. The factorization and renormalization scale
dependence in next-to-leading order is reduced from typically $10\%$
to $5\%$ or less. The variation of both scales simultaneously results
in strong cancellations. For the exclusive $\tilde{t}+$jet production
we have resummed the leading large logarithms for small transverse
momentum. The matched distributions exhibit some numerical dependence
on the matching; especially for large stop masses at the LHC the
resummation significantly enhances the differential cross sections up
to a matching point $p_T^{\rm match} \sim m/3$.  Finally, we note that
if we had a way to predict the helicity structure of the coupling
$\lambda''$, this process would currently be the only way to directly
determine the stop mixing angle at a hadron
collider~\cite{stop-pair}.\bigskip

%%%%%%%%%%%%%%%%%%%%  ACKNOWLEDGMENTS  %%%%%%%%%%%%%%%%%%%%

\noindent 
{\bf Acknowledgments}
\\[1mm]
We would like to thank T.~Han, M.~Spira and W.~Beenakker for very
helpful discussions and T.~Falk for carefully reading the manuscript.
Furthermore, we want to acknowlege the very pleasant discussions with
B.~Harris, Z.~Sullivan and E.L.~Berger, the authors of the original
leading order analysis~\cite{bhs}, which also inspired the calculation
of the exclusive cross sections for small and large transverse
momenta.

%%%%%%%%%%%%%%%%%%%%%%%  REFERENCES  %%%%%%%%%%%%%%%%%%%%%%%

\bibliographystyle{plain}

\end{document}